\documentclass{vgtc}                          

\usepackage{graphicx}
\usepackage{times}
\usepackage{mathptmx}

\onlineid{0}

\vgtccategory{Research}

\vgtcinsertpkg

\title{Rotation Blurring: Use of Artificial Blurring to Reduce Cybersickness in Virtual Reality First Person Shooters}

\author{Pulkit Budhiraja\thanks{e-mail:budhirj2@uiuc.edu} %
\and Mark Roman Miller\thanks{e-mail:mrmillr3@uiuc.edu} %
\and Abhishek K Modi\thanks{e-mail:akmodi2@uiuc.edu} %
\and David Forsyth\thanks{e-mail:daf@uiuc.edu} %
\affiliation{\scriptsize University of Illinois at Urbana-Champaign, Urbana, Illinois, United States}}

\abstract{
Users of Virtual Reality (VR) systems often experience vection, the perception of self-motion in the absence of any physical movement. While vection helps to improve presence in VR, it often leads to a form of motion sickness called cybersickness. Cybersickness is a major deterrent to large scale adoption of VR.

Prior work has discovered that changing vection (changing the perceived speed or moving direction) causes more severe cybersickness than steady vection (walking at a constant speed or in a constant direction). Based on this idea, we try to reduce the cybersickness caused by character movements in a First Person Shooter (FPS) game in VR. We propose Rotation Blurring (RB), uniformly blurring the screen during rotational movements to reduce cybersickness. We performed a user study to evaluate the impact of RB in reducing cybersickness. We found that the blurring technique led to an overall reduction in sickness levels of the participants and delayed its onset. Participants who experienced acute levels of cybersickness benefited significantly from this technique.
} 

\CCScatlist{ 
  \CCScat{H.5.1}{Information Interfaces and Presentation}{Multimedia Information Systems}%
{Artificial, augmented, and virtual realities};
  
}

\begin{document}


\firstsection{Introduction}

\maketitle


Low cost Virtual Reality (VR) Head Mounted Displays (HMDs) are now easily available.
The stereoscopic display in HMDs creates 3D Virtual Environment (VE) with large field of view and an enhanced sense of presence. But users often experience vection while exploring these VEs. Vection is the perception of self-motion caused by viewing a moving visual stimulus while not actually moving (eg the illusion of motion obtained in a stationary train watching an adjacent train move). 
Moving through in a virtual environment adds to the sense of presence in that environment~\cite{riecke2010compelling}, but often leads to a form of motion sickness called cybersickness, which often accompanies vection. Symptoms include dizziness,  fatigue, cold sweat, oculomotor disturbances, disorientation, nausea and (rarely) vomiting. Oculomotor disturbances and disorientation are more common with cybersickness than motion sickness~\cite{lawson2014handbook1}. Long exposures to VR exacerbates cybersickness, while repeated exposures to VR reduces the severity and incidence of cybersickness~\cite{kennedy2000duration}.

Our objective in this work was to reduce cybersickness experienced in VR gaming to improve comfort and enjoyability in a way that required no additional effort from game developers. We therefore set the constraint on our solution to be that it should have insignificant computational load and work solely with the rendering pipeline.

Prior work has found that vection that changes (through changes in perceived speed or direction) induces more severe Simulator Sickness (SS) than steady vection (walking at a constant speed or in a constant direction)~\cite{Bonato2008}. Although minor differences exist between cybersickness and SS~\cite{stanney1997cybersickness}, they are closely related. Work by Trutoiu et al.~\cite{Trutoiu2009} suggests that among all forms of movement, rotation causes the maximum amount of SS, and so we focused our attention on reducing cybersickness during rotations. Riecke et al.~\cite{riecke2006cognitive} show that photorealistic looking virtual environments enhance the amount of vection users experience when compared to an abstract version of the same virtual environment.

In this paper, we describe a novel navigation technique for VR games called Rotation Blurring (RB) which can help to reduce cybersickness. Our technique blurs the rotational movements triggered by an external controller in the virtual world. We hypothesize that blurring the rotations will make the parts that cause the most cybersickness look less photorealistic, thereby suppressing the overall level of cybersickness induced by those movements. We conducted a user study to evaluate this hypothesis and found that the technique helped users who are sensitive to cybersickness. For the user study, we chose to test our technique in a First Person Shooter (FPS) game. The high action gameplay of the FPS requiring continuous navigation makes it an ideal environment to test our technique.

\section{Related Work}

One of the first works to explore the impact of vection on SS by Hettinger et al.~\cite{hettinger1990vection} exposed users to a fixed-based flight simulator and measured SS and vection levels. One theory for cybersickness induced by vection is the mismatch of motion information from the visual system and the vestibular system during vection leading to cybersickness~\cite{oman1990motion}. Another theory posits that changes in stability of the human balance mechanism causes cybersickness ~\cite{riccio1991ecological}. Keshavarz et al.~\cite{Keshavarz2015} provide an in-depth review of works in this area.

Cybersickness is a major deterrent to large scale adoption of VR devices and researchers have explored different ways to reduce it. Dorado and Figueroa propose that ramps induce less cybersickness than stairs in VR~\cite{Dorado2014}. Domeyer et al. studied the effects of breaks between consecutive driving simulator sessions on reducing SS for older drivers~\cite{domeyer2013use}. Jeng-Weei Lin et al. discovered that providing motion prediction cues in driving simulators helps to reduce SS while not affecting presence~\cite{jeng2005unobtrusive}. Adding a virtual guiding avatar that provides motion cues helps to reduce SS in driving simulators~\cite{lin2004virtual}; the avatar also enhances a user's sense of presence in the virtual environment. Carnegie et al.~\cite{carnegie2015reducing} try to reduce visual discomfort by adding depth of field blur to the virtual scene. Leroy et al.~\cite{leroy2012real} describe an algorithm for implementing a real-time adaptive blur to remove irritating high frequency content in high horizontal disparity zones of stereoscopic display, so reducing eye strain caused by stereoscopic displays. Jung et al.~\cite{jung2013visual} describe a selective depth of focus blur technique that is applied only to regions that induce high visual discomfort but are less important visually. Blum et al. investigated the effectiveness of adding artificial out of focus blur on visual discomfort to increase fusion limits of double vision occurring in stereoscopic displays~\cite{Blum2010}.  In contrast to these works that use blur to reduce cybersickness, our approach uses blurring to minimize cybersickness arising particularly due to locomotion in VR. Furthermore we blur the whole screen as opposed to selectively blurring parts of the virtual environment.

\begin{figure*}[tbh]
\vspace{-8pt}
	\centering
	\includegraphics[width=7in]{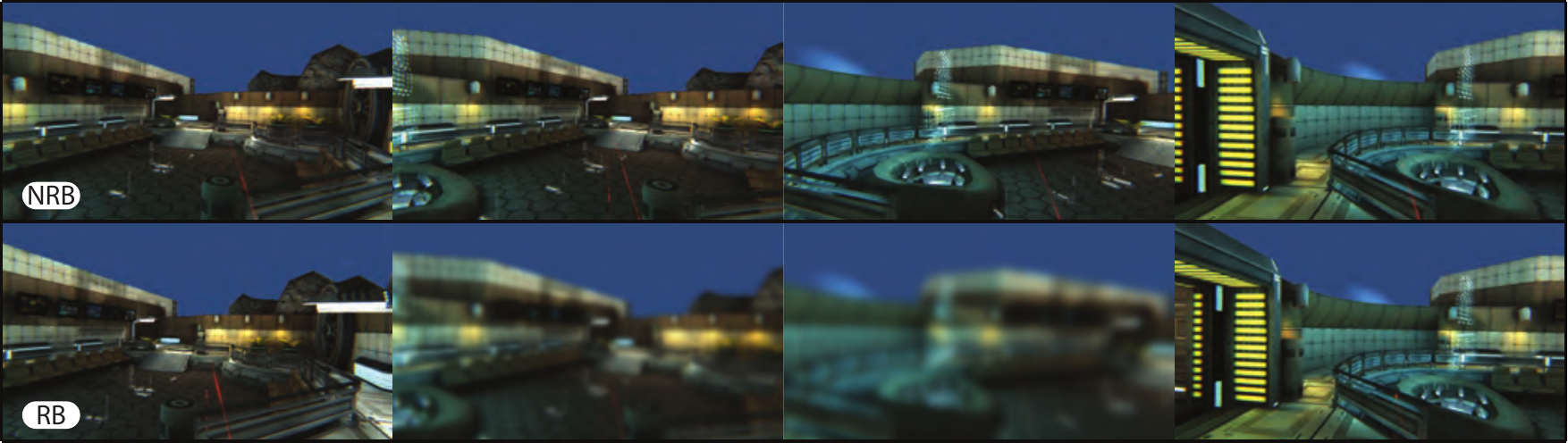}
	\caption{ Screenshots of 2 versions of the game during rotational movement. One with Rotation Blurring disabled (NRB) and one with Rotation Blurring enabled (RB).}
	\label{fig:comparison}
	\vspace{-15pt}
\end{figure*}

\section{Blurring Technique Description}

Rotation Blurring (RB) is implemented using shaders in Unity3D, a popular game engine that supports VR headsets. Since most FPS games use mouse controller to rotate the game character, we use the movement of the mouse as primary input signal to trigger and control RB. Our technique applies a uniform Gaussian blur to the screen whenever the mouse is used to rotate the game character. The amount of blur (standard deviation of the Gaussian function) is directly proportional to the magnitude of acceleration of the mouse movement. This proportionality ensures a smooth transition from a non-blurred to a blurred screen (and vice versa) as opposed to discrete jarring jumps between non-blurred and blurred screen.

RB is applied such that it only activates for rotations when navigating the game environment and not during other small movements like aiming the gun or dodging enemy fire. This is ensured by a combination of two rules. First, the blurring only activates when the magnitude of acceleration of the mouse is over a certain minimum threshold value. Second, the blurring would only happen if a continuous mouse motion is detected for a time threshold of 5 frames.

For this research, we used an Oculus Rift DK2 HMD which tracks the orientation and position of the user's head. Our technique aims to reduce cybersickness occuring from vection during rotational movements. Since cybersickness from vection only triggers in the absence of actual physical movements, any changes in the view point caused by user's real world head movements were not blurred.

\section{User Study}

In order to understand the effects of RB on cybersickness, we conducted a within-subjects user study where participants played a VR FPS game which was modified to enable RB. For comparison, participants also played another version of the game with RB disabled. We hypothesized that adding RB to the game will significantly reduce cybersickness experienced by the participants. 

For this study, we used an Oculus Rift DK2 HMD connected to a Windows PC. To gauge the sickness levels of participants, we used the standard 16 question Simulator Sickness Questionnaire (SSQ)~\cite{kennedy1993simulator} where each question could be answered on a Likert scale of 0-3. Responses of the SSQ were used to derive Total Sickness (TS) levels of the participants in the study.

\subsection{Virtual Scene Description}


We created the FPS shooter game using an open source Unity3D project called AngryBots. The game is set in an industrial setting and features many enemy bots spread across the arena. Participants used keyboard and mouse to control the character (using standard FPS controller layout). We created two versions of the game - with and without RB. We refer to these as RB and NRB respectively (Figure~\ref{fig:comparison}). 

In the game, the view vector of the character was coupled with the aim vector of the gun. Participants controlled the aim vector of the gun using the mouse. The coupling ensured that users did not have to move their head in conjunction with the mouse movement in order to see the gun's pointer. However, user's head movements could still independently control the viewing vector of the character. Doing this ensured that users did not have to move their head in the game all the time and suffer from the resulting neck sprain. 

In order to reduce cybersickness from other types of movements except rotation, some modifications were made to the character controls. Rotational movements with the mouse were only restricted in the horizontal plane. This was done since past research claims that frequent rotations with two degrees of freedom cause significantly more SS than rotations with one degree of freedom~\cite{bonato2009combined}. The aim vector was fixed to be parallel to the ground and the game was modified to ensure all the enemy bots could be destroyed without any vertical rotations.Work by Trutoiu et al.~\cite{Trutoiu2009} suggests that strafing (linear movement in left/right direction) is the most unconvincing form of movement in VR and also the second most SS causing form of movement after rotation. Hence, strafing movements in the game were disabled. In order to minimize cybersickness from linear walking, the character moved at a constant speed~\cite{Bonato2008}. 

\subsection{Procedure}

The study was conducted over two sessions on consecutive days. During each session, participants were tasked to play either the RB version or the NRB version of the game. The ordering of the tasks was determined randomly. We avoided performing both tasks on the same day to avoid the cumulative nature of nausea from biasing the amount of cybersickness experienced by participants in the latter task. Before each task, participants were pre-screened for good health by measuring their TS levels determined from a SSQ filled before each session. Any user with a TS levels of over 7.48 was rejected for the study as recommended in~\cite{lawson2014handbook2}. This ensured that participants were in an equally healthy mental state before both sessions. All participants had normal or corrected to normal vision. Any participant who used glasses were given the option to use them inside the headset.

In each session, users played the game for 10 minutes with the objective of destroying most number of bots. During the game, after every 2 minutes, participants were shown a sickness scale from 0-6 and were asked to verbally report their level of motion sickness symptoms. We used the motion sickness rating scale used in~\cite{griffin2004visual} (which is a minor modification of the scale from~\cite{golding1992comparison}). We advise readers to refer to figure~\ref{fig:nausearate} for a detailed description of each rating level.

At the end of the task, users were asked to fill the SSQ to gauge their post-task sickness levels. This was followed by a post study questionnaire and an interview. 

We received 18 responses from interested individuals for the study (14 males, 4 females) from an age group of 18 to 26 years. Users were required to have some past experience of playing FPS games. 1 user was rejected during pre-screening process as his/her pre-study TS levels was higher than acceptable limit. 2 other users got very sick within first 2 minutes of the study and could not continue. In the end, we had 15 participants (12 males, 3 females) successfully complete the study.

\subsection{Results}

Participants were asked to fill a SSQ before and after each task. The pre-task SSQ response was used for pre-screening participants for good health before the start of a task. Participants' post task questionnaire responses were used to derive their TS levels for the task. 

\begin{figure}[h]
	\vspace{-8pt}
	\centering
	\includegraphics[width=3.35in]{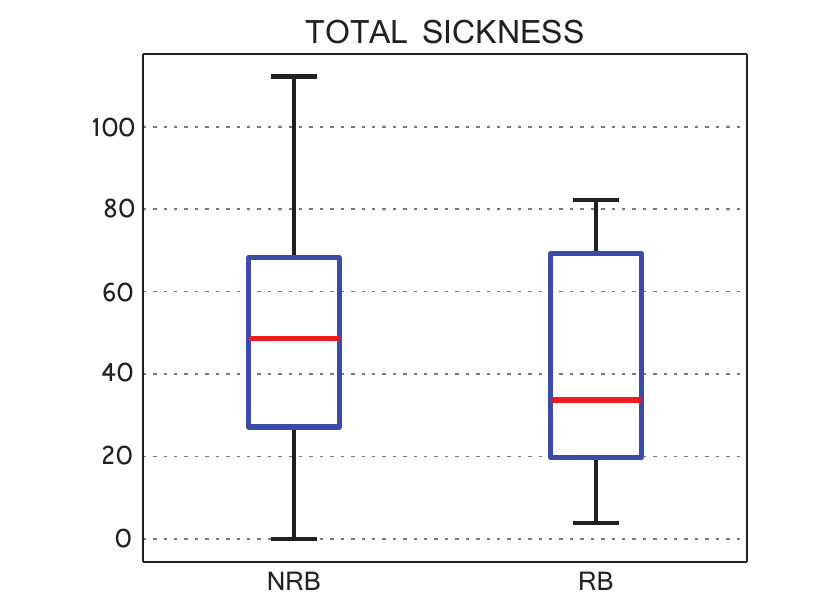}

	\caption{ Boxplot of Total Sickness levels obtained from SSQ. $mean^{NRB}_{TS} = 51.36,\ mean^{RB}_{TS} = 42.14,\ median^{NRB}_{TS} = 48.62,\ median^{RB}_{TS} = 33.66$.}
	\label{fig:boxplotTS}
	\vspace{-8pt}
\end{figure}

\begin{figure}[h]
\vspace{-8pt}
	\centering
	\includegraphics[width=3.35in]{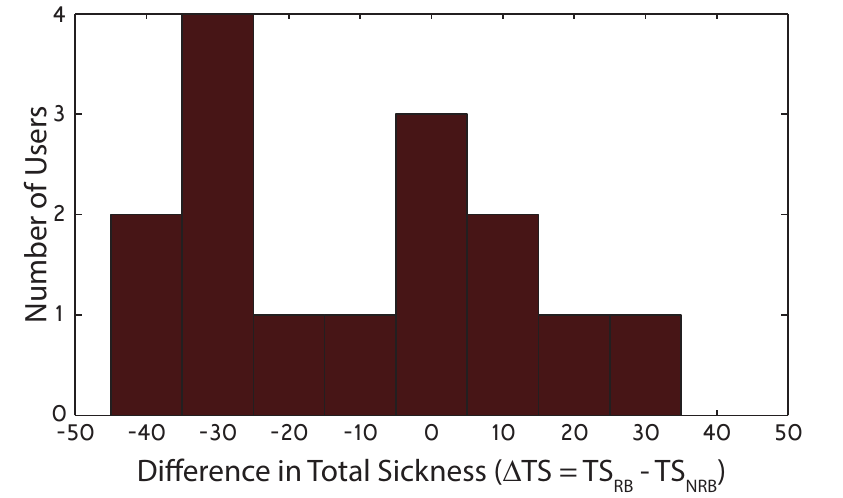}
	\vspace{-12pt}
	\caption{ Histogram of $\Delta$TS. On enabling RB, 8 participants experienced a decline, 3 experienced no change and 4 experienced an increase in cybersickness. The decline was substantial for 6 out of 8 participants.}
	\label{fig:histogram}
	\vspace{-8pt}
\end{figure}

Figure~\ref{fig:boxplotTS} shows the aggregated TS results for the NRB ($\mu = 51.36, \sigma = 34.67$) and RB ($\mu = 42.14, \sigma  = 27.61$) conditions. The mean TS response on enabling RB went down from 51.36 to 42.14 and the median values went down from 48.62 to 33.66. We performed a Wilcoxon signed-rank test to test the statistical significance of decrease in TS and found it to be not statistically significant ($p = 0.19$). Observations from figure \ref{fig:histogram} give an insight for this. 
	
Figure \ref{fig:histogram} shows a histogram of difference in TS levels when RB is enabled ($\Delta$TS = TS$_{RB}$ - TS$_{NRB}$). Out of 15 participants, 8 saw a decline in TS on enabling RB while 3 saw no change and 4 participants saw an increase in TS. Group of participants whose TS levels decreased benefited from RB while the other group did not. For sake of convenience, we'll refer to these groups as G1 and G2 respectively. It is interesting to observe the trends for the two groups. While a majority of participants in G2 saw no change or mild increase in TS levels ($\mu^{G2}_{\Delta TS} = 12.82$), the majority in G1 saw a significant drop in their TS levels ($\mu^{G1}_{\Delta TS} = -28.51$) suggesting that the ones who found RB to be beneficial benefited significantly from it. Further evaluation of the data reveals that the participants of G1 were more sensitive to cybersickness than those of G2 (TS levels without RB (NRB) - $\mu^{G2}_{TS} = 31.52$ vs $\mu^{G1}_{TS} = 68.72$). The results indicate that users who are sensitive to cybersickness are more likely to benefit from RB.

The subjective opinions of participants recorded during the interview about RB shows the similar bipolar trend. Most participants found blurring to be helpful and it did not bother them. One participant said \emph{``I think blur helped. It was jarring at first but I got used to it. I do think it helped''}. Another participant responded \emph{``Movement being blurred made it much easier. My eyes didn't get dizzy this time. I was scared last time to make quick turns (NRB). I feel I could go on for 2 more hours!''}. Some users disliked blurring. One participant commented \emph{``blurring distracted me from gameplay''}. Another participant said \emph{``Blurring made it harder to connect. Caused discomfort''}.

\begin{figure}[h]
\vspace{-8pt}
	\centering
	
	\includegraphics[width=3.35in]{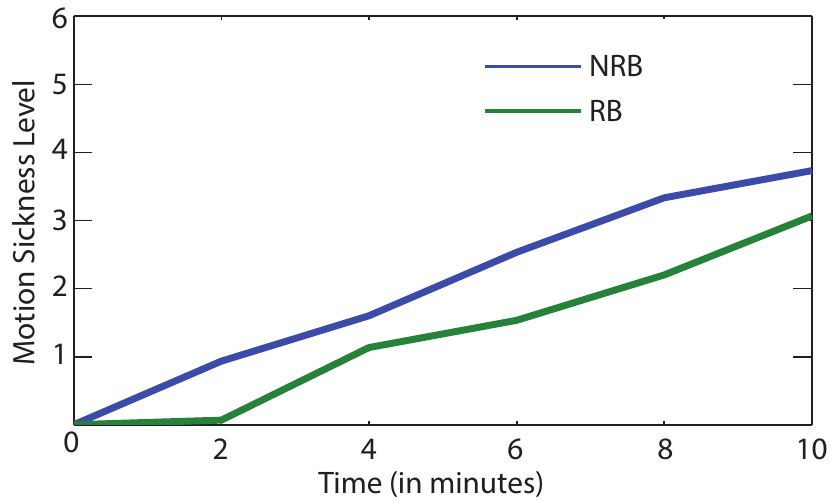}
	\vspace{-12pt}
	\caption{ Plot of mean motion sickness levels vs time on a scale of 0-6. 0 - no symptoms, 1 - any unpleasant symptoms, 2 - mild unpleasant symptoms, 3 - mild nausea, 4 - mild to moderate nausea, 5 - moderate nausea but can continue and 6 - moderate nausea, want to stop \vspace{-8pt}}
	\label{fig:nausearate}
		\vspace{-4pt}
\end{figure}

Participants were instructed to report their level of motion sickness symptoms on a scale of 0-6 after every 2 mins of gameplay. Figure~\ref{fig:nausearate} shows a plot of mean motion sickness levels at 2 minute intervals for NRB and RB versions of the game. Fig ~\ref{fig:nausearate} indicates that blurring helped delay the onset of cybersickness.

\begin{figure}[h]
\vspace{-8pt}
	\centering
	\includegraphics[width=3.35in]{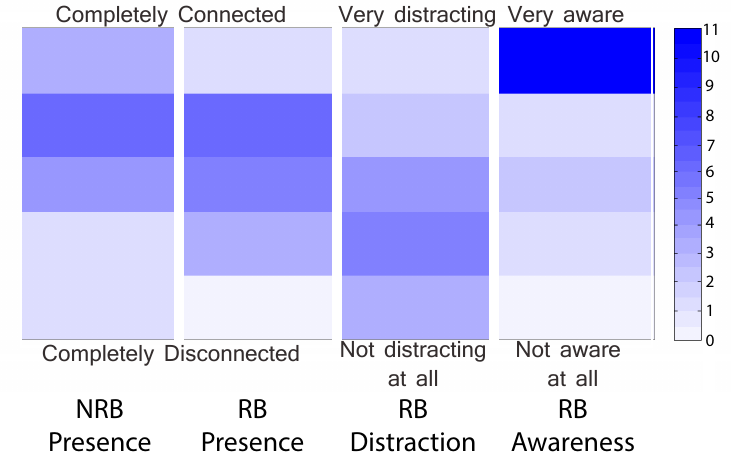}
	\caption{ Post Study Questionnaire Responses \vspace{-8pt}}
	\label{fig:questionnaire}
	\vspace{-8pt}
\end{figure}

Participants filled out a post study questionnaire to rate awareness of RB, distraction caused from RB and presence in VR, with and without RB, on a 5 point Likert scale. Figure \ref{fig:questionnaire} shows a summary of those responses. As the figure indicates, participants were highly aware of RB and did not find RB very distracting. RB did not significantly decrease their level of presence in VR.

\section {Discussion and Future Work}

The results from the user study show that RB helped in decreasing the overall cybersickness experienced by the participants. While it did not help some of the participants, it significantly helped those who experienced acute level of cybersickness. Post study interview comments about preference for RB also show this trend. The results show the positive impact of RB on delaying the onset of nausea. All participants saw a gentler growth in nausea level over time when RB was enabled.


These results open up the opportunity to conduct further research in exploring other ways of applying blur to cybersickness inducing movements in VR. Since users are more perceptive to motion in the peripheral regions of the eye, future work can explore the effect of applying only peripheral blur during rotations to reduce cybersickness. Other image effects like mosaic effects, edge segmentation and grayscale coloring that reduce the photorealism of the scene can be explored to understand their impact on reducing cybersickness.

Since some participants found the blurring to be ineffective and disruptive, future Just-Noticeable Differences (JND) studies can find the optimal response curve of blurring relative to movement in the game and the optimal amount of blur to add to the game which minimizes cybersickness while minimizing discomfort.

Preliminary results indicate minimal impact on presence by RB. While these results look promising, future work needs to explore this in an in-depth manner using the presence questionnaire to get a better understanding of the impact of RB on presence in VR. 

Our technique requires getting input from external controllers and applying image effects shaders, both of which can be done without access to the game level code. Being independent from the game, RB can be implemented by HMD developers at the SDK level as an optional setting to existing games without requiring original game developers to make any changes. Users who benefit from RB can enable it in games and have a better VR gaming experience while others can keep it off.

\section {Conclusion}


The results of the user study indicate that the Rotation Blurring (RB) technique helped to reduce mean cybersickness levels of the sample space. The results are not statistically significant since a set of participants did not benefit from the technique or got affected negatively. However, the set of users who experienced acute levels of cybersickness saw sharp falls in their levels of cybersickness on enabling RB and can largely benefit from this technique. Results also indicate that RB helped delay onset of cybersickness that can lead to longer gameplay periods in VR.

These results open up the opportunity to explore the impact of RB on a bigger sample set of users who are prone to cybersickness.


\bibliographystyle{abbrv}
\bibliography{v0.1}
\end{document}